\documentclass[twocolumn,amsmath,amssymb]{revtex4}
\usepackage{graphicx}
\usepackage{dcolumn}
\usepackage{bm}
\begin{document}

\title{A Stable Pfaffian State in Bilayer Graphene}
\author{Vadim M. Apalkov}
\affiliation{Department of Physics and Astronomy, Georgia State University,
Atlanta, Georgia 30303, USA}
\author{Tapash Chakraborty$^\ddag$}
\affiliation{Department of Physics and Astronomy,
University of Manitoba, Winnipeg, Canada R3T 2N2}

\date{\today}
\begin{abstract}
Here we show that the Pfaffian state proposed for the $\frac52$
fractional quantum Hall states in conventional two-dimensional
electron systems can be readily realized in a bilayer graphene
at one of the Landau levels. The properties and stability of
the Pfaffian state at this special Landau level strongly depend
on the magnetic field strength. The graphene system shows a
transition from the incompressible to a compressible state with
increasing magnetic field. At a finite magnetic field
of $\sim10$ Tesla, the Pfaffian state in bilayer graphene
becomes more stable than its counterpart in conventional electron systems.
\end{abstract}
\maketitle
Ever since the discovery of the quantum Hall state at the Landau
level filling factor $\nu=\frac52$, the first even-denominator
state observed in a single-layer system, it has been very
aptly characterized as an ``enigma" \cite{jim_review}. It was
clear at the outset that this state must be different from the
fractional quantum Hall effect (FQHE) in predominantly odd-denominator
filling fractions \cite{stormer,fqhe_book}. Understanding this
enigmatic state has been a major challenge in all these years
\cite{jim_tilted}. At this half-filled first excited Landau level
\cite{note1}, a novel state described by a pair wave function
involving a Pfaffian \cite{read,greiter} (or anti-Pfaffian
\cite{anti}) has been the strongest candidate. More intriguing 
are the elementary charged excitations at this ground state that 
have a charge $e^*=e/4$ and obey `non-abelian' statistics
\cite{halperin,stern}. Recent observation of the $e^*=e/4$
quasiparticle charge at $\nu=\frac52$ quantum Hall state
\cite{heiblum} has brought the issue to the fore \cite{storni}.
It has been suggested that these non-abelian quasiparticles,
besides carrying the signatures of Majorana fermions \cite{majorana}
in this system, might even be useful for quantum information
storage and processing in an intrinsically fault-tolerant manner
\cite{kitaev}.

Electrons in another recently discovered two-dimensional system,
graphene \cite{novo}, display a range of truly remarkable behavior
\cite{review}. The dynamics of electrons in a single sheet of
graphene, a hexagonal honeycombed lattice of carbon atoms is that
of massless Dirac fermions with linear dispersion, chiral
eigenstates, valley degeneracy, and unusual Landau levels in an
external magnetic field \cite{review}. Theoretical studies of
FQHE in monolayer \cite{fqhe_mono} and bilayer graphene
\cite{fqhe_bilayer} were reported earlier by us. Recent experimental
observations of the $\nu=\frac13$ FQHE in monolayer graphene
\cite{fqhe_graphene} have provided a glimpse of the role highly
correlated electrons play in graphene. Given the accute interest in
studying the properties of the $\nu=\frac52$ state in conventional
two-dimensional electron gas (2DEG), a natural question to ask is how
does this state manifests itself in graphene. 

For the conventional ({\it nonrelativistic}) 2DEG the incompressible 
state at $\nu=\frac52$ has been studied numerically for a finite number 
of electrons \cite{storni}. A relatively good (but not 100\%) overlap 
with the Pfaffian state has been found. The overlap of the exact 
wave function of the finite-size systems with the Pfaffian state can be 
improved by varying the inter-electron potential. For example, by increasing 
the thickness of the two-dimensional layer \cite{peterson}, one can 
improve the overlap with the Pfaffian state and increase the excitation 
gap of the corresponding incompressible state. The interaction properties 
of a two-dimensional system are determined by the Haldane pseudopotentials
\cite{haldane}, which are the energies of two electrons with 
relative angular momentum $m$. The pseudopotentials at the $n$-th Landau level are of the form

\begin{equation*}
V_m^{(n)}=\int_0^{\infty}\frac{q dq}{2\pi} V(q)
\left[F_{n}(q)\right]^2 L_m(q^2)e^{-q^2},
\end{equation*}
where $L_m(x)$ are the Laguerre polynomials,$V(q) = 2\pi e^2/(\kappa \ell^{}_0 q)$ is the Coulomb interaction potential
in the momentum space, $\kappa$ is the dielectric constant, $\ell^{}_0 =(\hbar/eB)^{\frac12}$ is the magnetic length,
and $F_{n}(q)$ are the form factors of the $n$-th Landau level.

Within the framework of the Haldane pseudopotentials it is convenient
to study the finite-size system numerically in the spherical geometry.
The size of the sphere and the number of single-particle states are
determined by the parameter $S$, where $2S$ is the number of magnetic 
fluxes through the sphere in units of the flux quanta. The single-electron 
states are characterized by the angular momentum $S$, and its $z$ 
component $S_z$. For the many-electron system the corresponding states 
are classified by the total angular momentum $L$ and its $z$ component
\cite{fano}. For a system with $N$ electrons the $\nu=\frac12$ Pfaffian 
state is realized at $2S=2N-3$. Here the filling factor $\nu=\frac12$ 
is defined as the filling factor of a given Landau level. In spherical 
geometry the $\nu =\frac 12$ Pfaffian state is the exact ground state 
only for a very special type of three-particle interaction \cite{greiter}
when the three-particle interaction potential is non-zero only if the total
angular momentum of three particles is $3S-3$. For any two-particle 
interaction the $\nu=\frac12$ Pfaffian state is not an exact eigenstate, 
which makes it impossible to continuously connect the Pfaffian state 
to any exact eigenstate of the two-particle Hamiltonian. By varying 
the interaction function, i.e., the pseudopotentials, the close 
proximity to the Pfaffian function with an overlap of 99\% can be achieved. 
The $\nu=\frac12$ Pfaffian state is most sensitive to the lowest pseudopotentials, $V_1$, $V_3$, and $V_5$.

For a single graphene layer the Landau level wave functions are mixtures 
of those for Landau levels of nonrelativistic systems; for example the 
first Landau level in graphene can be expressed in terms of zero and the 
first Landau wave functions of the nonrelativistic system \cite{fqhe_mono}. 
As a result the form-factor in a single graphene layer takes the form 
$F_n(q)=[L_n(q^2/2)+L_{n-1}(q^2/2)]/2^{\frac12}$ for $n\geq 1$ and 
$F_{n=0}=L_0(q^2/2)$ \cite{fqhe_mono}. Numerical analysis of finite-size 
systems in a spherical geometry with up to 14 electrons shows that the 
largest excitation gap around 0.02 $e^2/(\kappa\ell^{}_0)$ occurs at the 
$n=2$ graphene Landau level \cite{unpub}. Although the excitation gap at 
a finite size system in this case is comparable to the $\nu=\frac52$
nonrelativistic system the overlap of the ground state 
with the Pfaffian state is less than 0.5 at all Landau levels \cite{unpub}. 
This fact shows that a single graphene layer does not have stable 
incompressible $\nu=\frac12$ Pfaffian states. Modification of the interaction 
potential can improve the formation of the $\nu=\frac12$ incompressible 
Pfaffian state. One such modification can be achieved in a bilayer graphene. 

We show here that bilayer graphene can indeed improve the stability 
of the $\nu=\frac12$ Pfaffian state in graphene. Namely, one of the bilayer 
Landau levels (for a given valley) has a stable $\nu=\frac12$ Pfaffian 
state, the properties of which can be controlled by a magnetic field. 
The maximum overlap of the finite system ground state with the 
corresponding Pfaffian state occurs at finite values of the magnetic 
field. The $\nu=\frac12$ incompressible state of a bilayer graphene is 
more stable than the corresponding state in a conventional two-dimensional system.

\begin{figure}
\begin{center}\includegraphics[width=7.0cm]{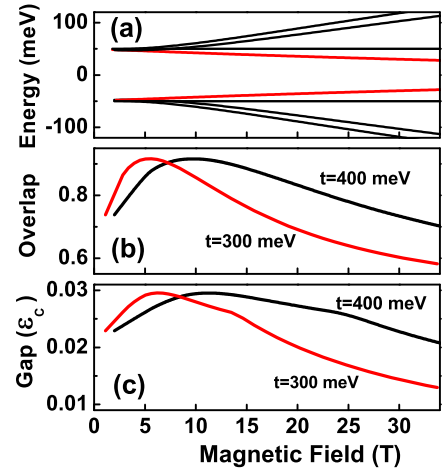}\end{center}
\vspace*{-0.5 cm}
\caption{(a) Few lowest Landau levels of a bilayer graphene,
shown for $\Delta U=100$ meV and $t=400$ meV. The two solid red lines 
belonging to different valleys, correspond to the Landau levels 
where the $\nu=\frac12$ Pfaffian state can be observed. 
(b) Overlap of the exact many-particle ground state with the 
Pfaffian function. (c) Excitation gap of the $\nu=\frac12$ 
state. The results are for $N=14$ and $2S=25$ and the zero bias 
voltage. The black and red lines correspond to $t=400$ meV and 300 meV, 
respectively. Here the energy unit is $\varepsilon^{}_c=e^2/\kappa\ell^{}_0$.}
\label{figone}
\end{figure}

We consider a bilayer graphene which consists of two coupled graphene
layers with the Bernal stacking arrangement. Each graphene layer has
two sublattices, say, A and B. For the Bernal stacking arrangement, the 
coupling is mainly between the atoms of sublattice A of the lower 
layer and atoms of sublattice B$^{\prime}$ of the upper layer. For 
one projection of spin, e.g., $+\frac12$, the state of the bilayer graphene 
can be expressed in terms of the four-component spinor $(\psi_A, \psi_B, 
\psi_{B^{\prime}}, \psi_{A^{\prime}})^T$ for valley $K$ and
$(\psi_{B^{\prime}}, \psi_{A^{\prime}}, \psi_A, \psi_B)^T$ for 
valley $K^{\prime }$. The subindices A, B and A$^{\prime}$, B$^{\prime}$ 
correspond to lower and upper layers respectively. The properties 
of bilayer graphene can be controlled by a bias voltage,
$\Delta U$, which is the potential difference between the upper 
and lower layers. The Hamiltonian of the biased bilayer system in a 
perpendicular magnetic field has the form \cite{pereira}

\begin{equation}
{\cal H} =  \xi \left(
\begin{array}{cccc}
    \Delta U/2 & v^{}_F \pi_+ & \xi t & 0   \\
    v^{}_F \pi_- & \Delta U/2 & 0 & 0 \\
    \xi t & 0& -\Delta U/2 & v^{}_F \pi_-   \\
    0 & 0&  v^{}_F \pi_+ & -\Delta U/2
\end{array}
\right),
\label{H1}
\end{equation}
where $t$ is the inter-layer hopping integral,
$\pi_{\pm} = \pi_x \pm  i \pi_y$,  $\vec{\pi} = \vec{p} + e\vec{A}/c$,
$\vec{p}$ is an electron two-dimensional momentum, $\vec{A}$
is the vector potential, $v^{}_F \approx 10^6$ m/s is the fermi velocity,
and $\xi=+$ ($K$ valley) or $-$ ($K^{\prime}$ valley).

\begin{figure}
\begin{center}\includegraphics[width=7.5cm]{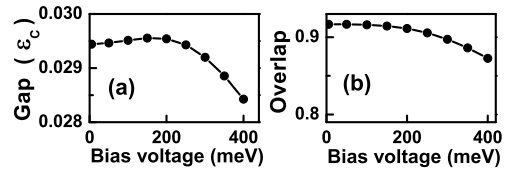}\end{center}
\vspace*{-0.5 cm}
\caption{The excitation gap (a) and the overlap with the Pfaffian state 
(b) versus the bias voltage, $\Delta U$. Here $N=14$ and $2S =25$, 
and the magnetic field is 10 Tesla.}
\label{figtwo}
\end{figure}

The discrete eigenstates of the Hamiltonian (\ref{H1}) can be 
found from the following equation \cite{pereira}
\begin{equation}
\left[\left(\varepsilon + \xi \delta\right)^2 - 2(n+1)\right]
\!\!\! \left[(\varepsilon - \xi \delta )^2 - 2n \right] = (\varepsilon ^2 - \delta ^2 )t^2,
\label{level1}
\end{equation}
where $\delta = \Delta U/2$ and all energies are expressed in units 
of $\hbar v^{}_F/\ell^{}_0$. For a given value of $n$ there are four 
bilayer Landau levels which are characterized by the index $n$ and the 
energy $\varepsilon$ of the level. The corresponding wave functions
can be expressed in terms of $n$,  $|n-1|$, and $n+1$ conventional
Landau wave functions \cite{pereira}. The resulting form factors 
$F_{n,\varepsilon} (q)$ were derived in \cite{fqhe_bilayer}.

The form factors and the corresponding pseudopotentials 
allow us to find the energy spectrum of a finite $N$-electron 
system in the spherical geometry \cite{fqhe_bilayer,haldane}.
We report our calculations for $N=8$, 10, and 14 electron systems. 
To determine the incompressibility of the system we calculated the 
excitation gap and the overlap of the ground state wave function 
with the Pfaffian function. We consider only one valley, for example, 
valley K. The results are similar for K$^\prime$.

\begin{figure}
\begin{center}\includegraphics[width=9.0cm]{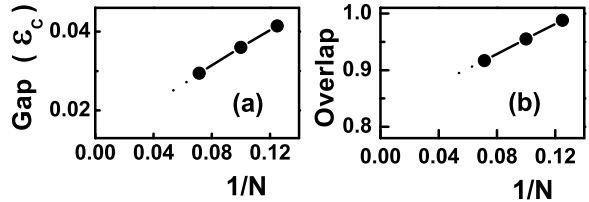}\end{center}
\vspace*{-0.5 cm}
\caption{The (a) excitation gap and the overlap with the Pfaffian 
state (b) are shown for different number of electrons: $N=8$, 10, 
and 14. The magnetic field is 10 Tesla and the bias voltage is zero. }
\label{figthree}
\end{figure}

For all but one bilayer Landau levels the overlap of the $\nu=\frac12$ 
ground state with the Pfaffian state is found to be small ($< 0.6$). 
At the same time there is one special Landau level (for each valley) 
at which the $\nu=\frac12$ ground state is well described by the Pfaffian 
function. This level corresponds to one of the solutions of 
Eq.~(\ref{level1}) with $n=0$. At a small bias voltage, $\Delta U $, the 
wave function corresponding to this Landau level is the mixture of the 
conventional Landau level wave functions with indices 0 and 1. The wave
functions of this special level have the form of  $(\phi_0,0,0,(t/\sqrt2)
\phi_1)$, where $\phi_n$ are $n$-th 'nonrelativistic' Landau 
functions and $t$ is in units of $\hbar v^{}_F/\ell^{}_0$. Then the 
corresponding form factor is  $F(q)=(L_0 + (t^2/2) L_1)/(1+t^2/2).$
At small values of the dimensionless hopping integral, $t(\ell^{}_0/
\hbar v^{}_F)$, the interaction within this level is similar to the 
one at the lowest Landau level of a conventional system, which does 
not show an incompressible $\nu=\frac12$ state. At large values of 
$t(\ell^{}_0/\hbar v^{}_F)$ the special bilayer Landau level 
is similar to the $n=1$ Landau level of the conventional system
and shows the $\nu=\frac12$ Pfaffian state. By varying the magnetic 
field, the dimensionless inter-layer hopping integral is changed
which modify the interaction within the Landau level and changes the 
properties of the $\nu=\frac12$ state. We present the numerical results 
only for this special bilayer Landau level.

\begin{figure}
\begin{center}\includegraphics[width=8.0cm]{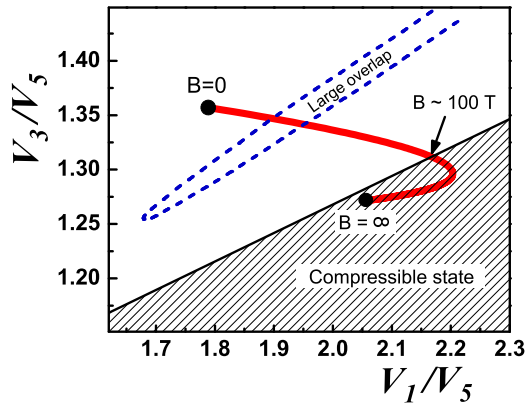}\end{center}
\vspace*{-0.5 cm}
\caption{Trajectory of the inter-electron interaction with varying 
magnetic field, shown by a solid red line in the plane $(V_1/V_5)-
(V_3/V_5)$ for the `special' Landau level of the bilayer graphene. 
The initial point of the trajectory (at $B=0$) corresponds to 
the nonrelativistic system at $n=1$ Landau level, while the final 
point (at $B=\infty$) corresponds to the nonrelativistic system at 
the $n=0$ Landau level. The shaded region illustrates the compressible 
$\nu=\frac12$ state, while the blank region corresponds to the
incompressible $\nu=\frac12$ state (Ref.~\cite{storni}). The crossing 
of the boundary between the compressible and incompressible states 
occurs at $B\sim 100$ Tesla for the hopping integral $t=400$ meV. 
The blue dashed line shows the region of large overlap with the
Pfaffian state (Ref.~\cite{storni}).}
\label{figfour}
\end{figure}

At the zero bias voltage this special Landau level has zero 
energy and is degenerate with another level, which has the form 
$(0,0,0,\phi_0)$. In addition to this accidental degeneracy, 
each level has two-fold valley degeneracy, which make the zero 
energy state four-fold degenerate. At a finite bias voltage this 
degeneracy is completely lifted and the special Landau level of 
the bilayer can be isolated. In Fig.~1(a) we show the lowest Landau 
levels of a bilayer at finite bias voltage. Two solid red lines 
correspond to the special Landau levels of the two valleys. 
The many-particles  properties of these two levels are identical. 
Therefore, we study the $\nu=\frac12$ state for only one valley. 

In Fig.~1(b,c) we show the magnetic field dependence of the overlap 
of the $\nu=\frac12$ ground state with the Pfaffian state and the 
corresponding excitation gap. At a small magnetic field the dimensionless 
hopping integral is large and the system becomes similar to the conventional 
system at the $n=1$ Landau level. With increasing magnetic field the 
properties of the system changes non-monotonically and the overlap 
with the Pfaffian state reaches its maximum at a magnetic field of
$\sim10$ Tesla (and for $t=400$ meV). The overlap at this point is 
$\approx 0.92$, which is a big improvement over the nonrelativistic 
system ($\sim 0.75$). The dimensionless hopping integral at this point 
is $t(\ell^{}_0/\hbar v^{}_F)\approx 4.89$. 

At a large magnetic field  the system is close to the $n=0$ 
nonrelativistic Landau level, the overlap with the Pfaffian 
state becomes small and the $\nu=\frac12$ state is finally  
compressible. This dependence on the magnetic field opens up
interesting possibilities to investigate the stability and 
appearance and disappearance of the $\nu=\frac12$ Pfaffian state 
in a single bilayer Landau level. Although the Pfaffian state 
becomes unstable only at large magnetic fields, this property 
strongly depends on the value of the hopping integral. At smaller 
hopping integrals the magnetic field range of stability of the 
Pfaffian state shrinks. For example, at $t=300$ meV the Pfaffian 
state is expected to be unstable at $B\sim 40$ Tesla (see Fig.~1). 
Another parameter which controls the properties of the graphene 
bilayer is the bias voltage. Although the bias voltage modifies 
the bilayer wave functions, we found that the overlap of the 
ground state with the Pfaffian state and the excitation gap 
have weak dependence on the bias voltage within a broad range 
of $\Delta U$ (see Fig.~2). The overlap monotonically decreases 
with increasing $\Delta U$, which suppresses the overlap by 
only a few percent. The large excitation gap and the large overlap 
observed for different system sizes are shown in Fig.~3.  

The stability and the strength of the Pfaffian state can be also
analyzed in terms of the general dependence of the pseudopotentials, 
$V_m$, on the relative angular momentum, $m$. We characterize the 
interaction potential of the bilayer graphene by two parameters: 
$V_1/V_5$ and $V_3/V_5$ \cite{storni}. These parameters depend 
on the magnetic field. By varying the magnetic field, this dependence 
can be shown as a line in the $(V_1/V_5)-(V_3/V_5)$ plane 
(Fig.~4). That line connects the initial point at $B=0$ to the final 
point, corresponding to large magnetic field, $B=\infty$. 
The $\nu=\frac12$ bilayer graphene system at the initial and 
final points are identical to the conventional systems at 
the first ($n=1$) and zero ($n=0$) Landau levels, respectively. 
In Ref.~\cite{storni} the region of the compressible $\nu=\frac12$ state 
and the region of strong overlap with the Pfaffian state were identified
(see Fig.~4). With increasing magnetic field, the $\nu=\frac12$ bilayer 
graphene system transforms from a $\nu=\frac52$ nonrelativistic state 
(at small values of $B$) to a more stable incompressible state with 
large overlap, and finally to a compressible state (at a large magnetic 
field). For the hopping integral $t=400$ meV, the transition from 
the incompressible to a compressible $\nu=\frac12$ state occurs at 
$B\sim 100$ Tesla. 

In conclusion, a stable incompressible $\nu=\frac12$ Pfaffian state 
can in fact be observed in a bilayer graphene only at one Landau level. 
The properties of this state strongly depend on the value of the magnetic 
field. With an increasing magnetic field, the $\nu=\frac12$ state 
transforms from an incompressible state at a small magnetic field to 
a compressible state at a large magnetic field. At intermediate values 
of the magnetic field, $B\sim 10$ Tesla, the $\nu=\frac12$ state 
becomes more stable than the corresponding state in a conventional 
two-dimensional electron system. 

The work has been supported by the Canada Research Chairs Program.


\begin{thebibliography}{99}
\bibitem{note1}
At this filling factor the lowest Landau level is fully occupied
with electrons of both spin polarizations. The remaining electrons
then fill half of the next Landau level.
\end{thebibliography}
\end{document}